\documentclass[sigconf]{acmart}
\AtBeginDocument{%
  }

\setcopyright{acmlicensed}
\copyrightyear{2018}
\acmYear{2018}
\acmDOI{XXXXXXX.XXXXXXX}

\acmConference[Conference acronym 'XX]{Make sure to enter the correct
  conference title from your rights confirmation email}{June 03--05,
  2018}{Woodstock, NY}
\acmISBN{978-1-4503-XXXX-X/2018/06}

\usepackage{multirow}

\begin{document}

\title{Token-level Collaborative Alignment for LLM-based Generative Recommendation}

\author{Fake Lin}
\email{fklin@mail.ustc.edu.cn}
\orcid{0009-0003-1402-2358}
\affiliation{%
  \institution{University of Science and Technology of China}
  \city{Hefei}
  \country{China}
}

\author{Binbin Hu}
\email{bin.hbb@antfin.com}
\orcid{0000-0002-2505-1619}
\affiliation{%
  \institution{Ant Group}
  \city{Hangzhou}
  \country{China}
}

\author{Zhi Zheng}
\email{zhengzhi97@ustc.edu.cn}
\orcid{0000-0001-7758-8904}
\authornotemark[1]
\affiliation{%
  \institution{University of Science and Technology of China}
  \city{Hefei}
  \country{China}}

\author{Xi Zhu}
\email{xi.zhu@rutgers.edu}
\orcid{0000-0003-3621-8493}
\affiliation{%
  \institution{Rutgers University}
  \city{New Brunswick}
  \country{United States}}

\author{Ziqi Liu}
\email{ziqiliu@antgroup.com}
\orcid{0000-0002-4112-3504}
\affiliation{%
  \institution{Ant Group}
  \city{Hangzhou}
  \country{China}
}

\author{Zhiqiang Zhang}
\email{lingyao.zzq@antfin.com}
\orcid{0000-0002-2321-7259}
\affiliation{%
  \institution{Ant Group}
  \city{Hangzhou}
  \country{China}
}

\author{Jun Zhou}
\email{jun.zhoujun@antfin.com}
\orcid{0000-0001-6033-6102}
\affiliation{%
  \institution{Ant Group}
  \city{Hangzhou}
  \country{China}
}

\author{Tong Xu}
\authornotemark[1]
\email{tongxu@ustc.edu.cn}
\orcid{0000-0003-4246-5386}
\affiliation{%
  \institution{University of Science and Technology of China}
  \city{Hefei}
  \country{China}}

\renewcommand{\shortauthors}{Trovato et al.}

\begin{abstract}
Large Language Models (LLMs) have demonstrated strong potential for generative recommendation by leveraging rich semantic knowledge. However, existing LLM-based recommender systems struggle to effectively incorporate collaborative filtering (CF) signals, due to a fundamental mismatch between item-level preference modeling in CF and token-level next-token prediction (NTP) optimization in LLMs. Prior approaches typically treat CF as contextual hints or representation bias, and resort to multi-stage training to reduce behavioral–semantic space discrepancies, leaving CF unable to explicitly regulate LLM generation. In this work, we propose Token-level Collaborative Alignment for Recommendation (TCA4Rec), a model-agnostic and plug-and-play framework that establishes an explicit optimization-level interface between CF supervision and LLM generation. TCA4Rec consists of (i) Collaborative Tokenizer, which projects raw item-level CF logits into token-level distributions aligned with the LLM token space, and (ii) Soft Label Alignment, which integrates these CF-informed distributions with one-hot supervision to optimize a soft NTP objective. This design preserves the generative nature of LLM training while enabling collaborative alignment with essential user preference of CF models. We highlight TCA4Rec is compatible with arbitrary traditional CF models and generalizes across a wide range of decoder-based LLM recommender architectures. Moreover, it provides an explicit mechanism to balance behavioral alignment and semantic fluency, yielding generative recommendations that are both accurate and controllable. Extensive experiments demonstrate that TCA4Rec consistently improves recommendation performance across a broad spectrum of CF models and LLM-based recommender systems. Our code is available at \href{https://github.com/critical88/TCA4Rec}{https://github.com/critical88/TCA4Rec}.
\end{abstract}



\received{20 February 2007}
\received[revised]{12 March 2009}
\received[accepted]{5 June 2009}

\maketitle
\section{Introduction}


With the advance of large-scale pre-training, Large Language Models (LLMs) \cite{bai2023qwen,zhao2023survey,wu2024survey, zhu2025llm} have demonstrated a strong capacity to encode rich and diverse world knowledge, providing semantic enhancement for user profiles and item attributes and thus revealing their great potential in recommender systems \cite{zhang2024notellm,zhu2025recommender,wang2025unleashing,DBLP:journals/corr/abs-2305-00447}.
Nevertheless, while LLMs excel at semantic modeling and reasoning, they struggle to capture the collaborative filtering (CF) structure embedded in user behavior sequences that traditional CF models explicitly exploit \cite{DBLP:journals/corr/abs-2305-00447,liao2024llara,DBLP:journals/corr/abs-2310-19488,DBLP:journals/corr/abs-2406-10450,liu2024end}. This structural tension between behavior modeling (CF) and knowledge modeling (LLMs) often forces models to compromise between semantic plausibility and behavioral consistency, emerging as a core challenge in existing LLM-based recommendations \cite{jiang2025beyond}.
Motivated by this observation, many efforts have been devoted to a fundamental question: how to effectively inject the essential CF signals into LLMs, enabling them to go beyond generic semantic understanding and better adapt to the recommendation task itself.

The first line of works injects CF signals into LLMs as soft prompts \cite{DBLP:journals/corr/abs-2310-19488,liao2024llara,lin2024rella,cui2024distillation}, e.g., by transforming user/item representations into LLM-compatible embeddings via adapter module  \cite{DBLP:journals/corr/abs-2310-19488,liao2024llara}. However, these methods not only suffer from the structural mismatch between behavioral and semantic spaces, but also resort to multi-stage training \cite{DBLP:journals/corr/abs-2310-19488} and curriculum training \cite{liao2024llara} to alleviate this gap, leading to reduced training controllability and increased computational overhead.
Alternatively, quantization-based approaches compress items into discrete token sequences as semantic IDs (SID) \cite{rajput2023recommender,zheng2024adapting,wang2024learnable}. While this strategy ensures semantic reconstruction, the resulting codebooks predominantly reflect semantic similarity among items, rather than collaborative structures that differentiate user preferences. Consequently, the desired user-specific and context-dependent preference patterns are substantially depressed, or even lost, during quantization, rendering them insufficient for aligning the CF signals required by the recommendation task. These limitations call for rethinking the role of CF signals: \textit{Should CF be passively embedded as semantic enhancer, or explicitly imposed as learning signals to actively shape model optimization? }

Along this line, we shift our attention to how CF signals should effectively guide the LLM generation. 
On the one hand, many approaches incorporate CF either as an contextual hints or as a representation-level inductive bias, where it primarily serves as a conditioning variable when conducting reasoning  \cite{liao2024llara}. When CF signals are excluded from the training objective, they can only passively affect model learning through forward propagation, leaving them unable to directly steer or correct the optimization trajectory \cite{wang2025beyond}.
More fundamentally, traditional CF models learn user preference through item-level ranking objectives, whereas LLMs are optimized under token-level next-token prediction (NTP). This inherent mismatch in optimization granularity and supervision form leaves no natural objective-level interface for collaborative alignment.
These observations motivate us to rethink the optimization coupling between CF and LLM generation: \textit{How can item-level preferences be effectively projected into a token-level objective, enabling CF signals to participate in LLM training and generation in a controllable manner?}

 In this paper, we present \textbf{Token-level Collaborative Alignment for Recommendation} (TCA4Rec), a model-agnostic and plug-and-play framework that establishes a principled token-level interface between CF signals and LLM generation for generative recommender systems. TCA4Rec consists of two tightly coupled components.
(1) Collaborative Tokenizer.
This module bridges the granularity gap between item-level CF supervision and token-level language modeling. Specifically, we extract raw CF logits from any pretrained CF model and transform them—via normalization and aggregation—into token-level distributions aligned with the LLM token space over valid candidate items. This transformation enables item-level preference signals to seamlessly align with token-level objectives, while maintaining information integrity.
(2) Soft Label Alignment.
Building upon the CF-informed token distributions, we construct soft labels by combining them with one-hot supervision, and optimize a soft next-token prediction (NTP) loss to guide LLM generation. This design preserves the intrinsic generative nature of LLMs, while explicitly injecting CF signals to reflect essential user preference, thereby enabling collaborative alignment with LLM generation.
Notably, TCA4Rec is model-agnostic and plug-and-play. It seamlessly integrates CF signals from arbitrary traditional CF models (e.g., SASRec, BERT4Rec), and is compatible with any decoder-based generative recommender architectures (e.g., TallRec, LLaRA, CoLLM, and MSL). Moreover, by operating directly at the token level, TCA4Rec provides an explicit mechanism to trade off collaborative alignment and semantic fluency, yielding controllable generative recommender systems.

In summary, the technical contributions of this paper could be summarized as follows:

\begin{itemize}
\item We present a novel perspective to impose CF signals to regulate LLM learning process, which effectively bridges the gap between CF models and LLMs.

\item We introduce a model-agnostic and plug-and-play framework TCA4Rec, which features a Collaborative Tokenizer for item-to-token CF projection and a Soft-label Alignment mechanism for collaborative-guided LLM optimization.
\item Extensive experiments are conducted on three public datasets to demonstrate that our proposed approach are superior over baselines and further validate that our method enhances alignment with CF signals.
\end{itemize}

\begin{table}[t]
\centering
\renewcommand{\arraystretch}{1.3}
\caption{Prompt Template for Generative Recommendation}
\label{tb:prompt}
\begin{tabular}{p{0.9\columnwidth}}
\hline
\textbf{Instruction} \\
\hline
Given the user interaction history, please recommend the toys that the user is most likely to view. \\
\hline
\textbf{Input} \\
\hline
User Interaction Histories: 
"Magna-Tiles ICE, 32 Piece Set", "Fisher-Price Octonauts Launch and Explore Octo-Lab", ...

Then the toys that user is most likely to view: \\
\hline
\textbf{Output} \\
\hline
Crayola 3CT Color Foam Toy
 \\
\hline
\end{tabular}
\end{table}

\section{Preliminaries}
\subsection{Sequential Recommendation Task}
Sequential recommendation task aims to predict the next item that a user will interact with based on their historical interaction sequence. Formally, let $\mathcal{U}$ and $\mathcal{I}$ denote the sets of users and items, where $|\mathcal{U}|=N_u$ and $|\mathcal{I}|=N_i$. For each user $u \in \mathcal{U}$, the interaction history is denoted as $H_u=\{i_1, i_2, \dots, i_k\}$. The goal is to train a recommendation model $f$ such that $f(u, H_u) = i_{k+1}$, meaning that the model can infer the next item $i_{k+1}$ from the user profile and its past interactions $H_u$. The formulation is as follows:
\begin{align}
   \mathcal{L}_{\text{Rec}}
= - \log P\!\left(i_{k+1} \mid u, H_u; \theta_{Rec}\right).
\end{align}
where $\theta_{Rec}$ means the parameters of sequential recommender systems.

\begin{figure*}
\centering
\includegraphics[width=\linewidth ]{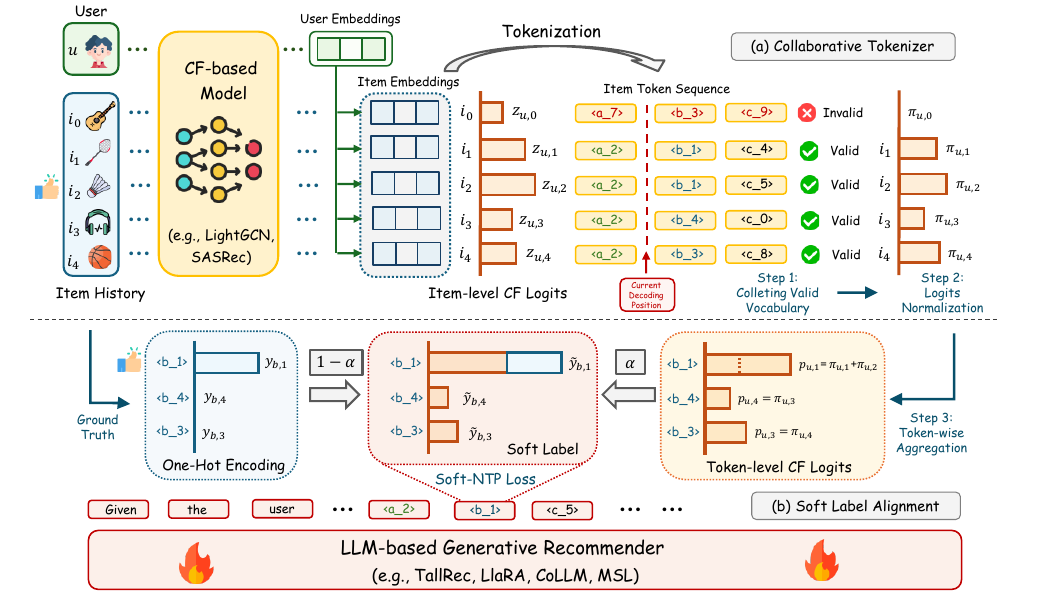} 
\caption{The overall framework of our method. (a) Collaborative Tokenizer module converts item-level logits from CF models into token-level distributions. (b) Soft Label Alignment module fuses these CF-informed token probabilities with one-hot supervision to construct soft labels. Finally, the LLM is fine-tuned with a soft next-token prediction loss which incorporates collaborative signals.}
\label{fig:framework}
\end{figure*} 
\subsection{LLM-based Recommendation Task}
To adapt sequential recommendation into the generative setting of LLMs, we reformulate it as a next item title prediction task. Suppose the interaction history of user $u$ can be represented as an item sequence $H_u = \{i_1, i_2, \dots, i_k\}$, where $i_k$ denotes the $k-th$ item in this sequence. 
We then construct a text prompt $x_u$ filled with metadata of items (e.g., title) from $H_u$. for clarity, we omit the subscript $i$m and then we arrive at the following derivation:
\begin{equation}
    \begin{aligned}
\label{eq:next_token_pred}
    \mathcal{L}_{NTP} &=-logP(t \mid x_u; \theta) = -\sum_{j=1}^{|t|}
\log P\!\left(y_{j} \mid x_u, y_{<j}; \theta\right) \\
    &= - \sum_{j=1}^{|t|}
\log \frac{exp(f_{\theta}(y_j \mid x_u,y_{<j}))}{\sum_{v\in \mathcal{V}} exp(f_{\theta}(v \mid x_u,y_{<j}))} \\
    &= - \sum_{j=1}^{|t|}
\log \frac{\sum_{v\in \mathcal{V}} \mathbf{1}_{v=y_j} \cdot exp(f_{\theta}(v \mid x_u,y_{<j})) }{\sum_{v\in \mathcal{V}} exp(f_{\theta}(v \mid x_u,y_{<j}))}.
\end{aligned}
\end{equation}
where $t=(y_1,\dots,y_{|t|})$ denotes the token sequence from target item $i$, $\mathcal{V}$ is the entire vocabulary of LLM, and $\theta$ are the LLM parameters.
Here $f_\theta(v \mid x_u, y_{<j})$ represents the LLMs that try to predict the next token $v$ given prompt $x_u$ and previous token sequence $y_{<j}$. 
$\mathbf{1}_{v=y_j}$ is the indicator that equals $1$ iff $v=y_j$ and $0$ otherwise. 
This objective follows the standard auto-regressive next-token prediction paradigm. We showcase a prompt template in the Table. \ref{tb:prompt}.
\section{Methodology}
In this section, we provide a detailed introduction to the TCA4Rec, which can be decomposed into two modules: (1) Collaborative Tokenizer, which transforms item-level logits from collaborative signals into token-level distributions. (2) Soft Label Alignment, which fuses them with one-hot labels to form soft labels and optimizes a soft next token prediction loss. The overall framework is illustrated in Figure~\ref{fig:framework}.

\subsection{Collaborative Tokenizer}
This module aims to transfer collaborative signals from the item-level logits to token-level distribution, making them compatible with language model training.
\subsubsection{Collaborative Logits Derivation}  Collaborative logits in recommendation models often capture user preferences for items, typically computed as the dot product between user and item embeddings. Thus, we first need to gather the embeddings from the recommendations. Here, ID-based sequential methods have shown remarkable power in the recommendation domain  \cite{DBLP:conf/sigir/0001DWLZ020,DBLP:conf/icdm/KangM18}. Following \cite{liao2024llara}, embeddings are obtained by the following formula:
\begin{equation}
\begin{aligned}
    e_{u} &= RS_{ID}(H_u; \theta^*_{ID}) \\
    e_{i} &= RS_{ID}(i; \theta^*_{ID})
\end{aligned}
\end{equation}
where $e_{u}$ and $e_{i}$ denotes the embeddings of user $u \in U$ and item $i \in I$, respectively. $H_u$ is the history interaction of users. $RS_{ID}$ is the related ID-based sequential recommender methods, such as SASRec \cite{DBLP:conf/icdm/KangM18}. And the $\theta^*_{ID}$ means the pretrained parameters of $RS_{ID}$. 
  
Then, we are supposed to obtain logits from these embeddings. 
We use scoring function to measure the relevance between the CF-based user embedding $e_u$ and the item embedding $e_i$:
\begin{align}
    z_{u} = \left\{\, z_{u,i} = g(e_u, e_i) \;\middle|\; i \in \mathcal{I} \,\right\}
\end{align}
where $g(\cdot, \cdot)$ denotes a general scoring function (e.g., dot product) that computes the logit between the user embedding $e_u$ and the item embedding $e_i$. It is worth noting that the choice of the scoring function $g(\cdot, \cdot)$ depends on the sequential recommendation model. For example, in SASRec, dot product is adopted to compute the score for ranking items by user. Therefore, we also employ dot product in this setting to ensure that the logits faithfully reflect the user’s preference over items. 

\subsubsection{Token-level Distribution Transformation}
Based on the item-level collaborative logits $z_u$ obtained in the previous subsection, we now investigate how to transform these logits into a token-level form. To achieve this goal, the first step is to translate item into sequence of tokens via tokenizer, such as LLM tokenizer for textual information. 
For each item $i \in \mathcal{I}$, it can be represented as a sequence of tokens:
\begin{equation}
t_i = (w_{i,1}, w_{i,2}, \dots, w_{i,m_i}),
\end{equation}
where $t_i$ means the token sequence of item $i$, $m_i$ denotes the length of $t_i$, and $w_{i,j}$ represents the $j$-th token of item $i$. Then, we adopt an intuitive grouping strategy to reorganize the CF logits and transform them into token-level probabilities. Specifically, this process is iteratively performed through three steps: 

\textbf{Collecting items.}
The purpose of this step is to determine which items are still valid candidates at the current decoding position. 
Recall that the LLM generates a sequence auto-regressively, that is, we predict $j-$token conditions on the prefix tokens $y_{<j}=\{y_1,\dots,y_{j-1}\}$. 
Therefore, we only consider those items whose token sequence begins with this prefix, since only these items remain consistent with the current generation. 
Formally, the candidate set is defined as
\begin{equation}
    \mathcal{C}(y_{<j})=\big\{\, i \in \mathcal{I}\;\big|\; j \le |t_i|,\ (w_{i,1},\dots,w_{i,j-1})=y_{<j}\,\big\},
\end{equation}
where $|t_i|$ is the sequence length, and $w_{i,k}$ denotes the $k$-th token of $t_i$.

\textbf{Normalizing logits.} Then we are supposed to transform the raw CF logits of candidate items into a valid probability distribution.
As logits are unnormalized scores, summing logits across different items would severely distort the original user preference towards these items. 
Therefore, we normalize them into probabilities via the softmax function, which ensures comparability and additivity across candidate items, making them suitable for subsequent aggregation at the token level.  
Formally, we apply the softmax restricted to $\mathcal{C}(y_{<j})$:
\begin{equation}
\pi^{(j)}_{u,i}\!\left(y_{<j}\right)
=\frac{\exp\!\big(z_{u,i}\big)}{\sum\limits_{k\in \mathcal{C}(y_{<j})}\exp\!\big(z_{u,k}\big)}\,
\end{equation}
Here, $\pi^{(j)}_{u,i}$ denotes the probability of item $i$ under the prefix $y_{<j}$.

\textbf{Aggregating probabilities.}
Since the LLM operates at the token level, we need to reorganize these item-level probabilities into token-level distributions.  
Concretely, we aggregate the probabilities of all items that share the same next token:
\begin{equation}
    p_u\!\left(v \mid y_{<j}\right)
=\sum_{i\in \mathcal{C}(y_{<j})} \pi^{(j)}_{u,i}\!\left(y_{<j}\right)\,\mathbf{1}\!\big[w_{j}=v\big]\, , \quad v \in \mathcal{V}
\end{equation}
where $v$ denotes the candidate token from LLM vocabulary at step $j$.  
In this way, the item-level logits are seamlessly transformed into token-level probability distributions at decoding position $j$, ensuring that the probabilities under each candidate token are properly normalized, i.e., \(\sum_{v\in \mathcal{V}} p_u(c\mid y_{<j})=1\).

After iteratively applying these three steps, we finally obtain the distribution of each token from the perspective of collaborative signals.

\subsection{Soft Label Alignment}
Although we have obtained the collaborative token-level distribution, we still have concerns about how to effectively incorporate this distribution into LLMs.
On the one hand, we need to preserve the intrinsic nature of next-token prediction, which is essential for maintaining the generative capability of LLMs.  
On the other hand, we aim to inject collaborative signals into the training objective so that the model can align its generative process with user–item interaction patterns.
To balance these two aspects, inspired by the label smoothing technique, we design a soft label alignment strategy, where the collaborative token-level distribution is fused with the original one-hot supervision. Formally, we extend the hard label, i.e., indicator $\mathbf{1}_{v=y_j}$ in Eq (\ref{eq:next_token_pred}), to arrive at the following soft label:
\begin{equation}
    \tilde{y}_j(v) = (1-\alpha)\,\mathbf{1}_{v=y_j} + \alpha\, p_u(v \mid y_{<j}), \quad v \in \mathcal{V}
\end{equation}
where $y_j$ is the ground-truth next token at decoding position $j$, $p_u(\cdot \mid y_{<j})$ is the collaborative token-level distribution, and $\alpha \in [0,1]$ is the hyper-parameter to control the balance between the two sources of supervision. 

This formulation indicates that during NTP training, the model should not only follow the strict ground-truth supervision but also take the collaborative knowledge into account. 
This enables the LLM to retain its generative nature while being gently regularized by user–item interaction patterns, which helps the model better capture collaborative semantics and improve its alignment with recommendation objectives.

Finally, by replacing the hard indicator $\mathbf{1}_{v=y_j}$ in Eq.~(\ref{eq:next_token_pred}) with the soft label $\tilde{y}_j(v)$, we obtain our final training objective, the soft cross-entropy loss:

\begin{equation}
    \begin{aligned}
    \label{eq:soft_label}
    \mathcal{L}_{\text{soft-NTP}}
    &= - \sum_{j=1} ^ {|t|} \log \sum_{v\in \mathcal{V}}\tilde{y}_j(v) \cdot P(v | x_u, y_{<j}; \theta) \\
    &= - \sum_{j=1}^{|t|} \log \frac{\sum_{v\in \mathcal{V}} \tilde{y}_j(v) \cdot \exp(f_{\theta}(v \mid x_u, y_{<j})) }{\sum_{v\in \mathcal{V}} \exp(f_{\theta}(v \mid x_u, y_{<j}))} \\
    \end{aligned}
\end{equation}
 This formulation degrades to the standard next-token cross-entropy loss when $\alpha=0$, and fully follows the collaborative distribution when $\alpha=1$. By adjusting $\alpha$, we can flexibly control the extent to which collaborative signals influence the generative learning objective. In this way, LLMs can fast adapt to the recommendation task and recognize the refined collaborative knowledge with minimal additional cost.

\subsection{Relation to Auxiliary KL Loss}
\label{sec:kl_loss}
\subsubsection{Auxiliary KL Formulation}
An alternative and plausible approach is to introduce a KL divergence as an auxiliary loss, which explicitly encourages the LLM’s predictive distribution to align with the collaborative distribution $p_u$. Concretely, we can define the following alternative objective:
\begin{equation}
\begin{aligned}
\label{eq:aux}
    \mathcal{L}_{\text{aux}}
&= (1-\alpha)\,\mathcal{L}_{\text{NTP}}
\;+\;
\alpha \sum_{j=1}^{|t|}
\mathrm{KL}\!\left(p_u(\cdot \mid y_{<j})
\,\|\, 
P(\cdot \mid x_u, y_{<j}; \theta)\right). \\
&= - \sum_{j=1}^{|t|}\sum_{v\in\mathcal V}
     \tilde{y}_j(v)\,\log P_j(v)
   + \text{const},
\end{aligned}
\end{equation}
For brevity, we denote the collaborative token-level distribution as 
$p_j(v)=p_u\left(v\mid y_{<j}\right)$, and the model prediction as $P_j(v)=P(v \mid x_u,y_{<j}; \theta)$ . This shorthand notation allows us to present the derivation more clearly. 
The specific derivation process can be found in Appendix \ref{form:aux_detail}.

\subsubsection{Gradient Analysis}
Although the two objectives appear similar at the surface level, their gradients with respect to the model parameters $\theta$ reveal a clear distinction. To explicitly observe this distinction, we analyze the gradient behaviors of the auxiliary KL formulation and the soft-NTP formulation  in Eq.~\ref{eq:soft_label}.

\textbf{Gradient of auxiliary KL.}
For Eq.~\eqref{eq:aux}, the gradient can be written as
\begin{equation}
\nabla_\theta \mathcal{L}_{\text{aux}}
= - \sum_{j=1}^{|t|}\sum_{v\in\mathcal V} \tilde{y}_j(v)\,\nabla_\theta \log P_j(v).
\end{equation}
When $P_j$ is parameterized by the softmax logits $z_j(v)$, this simplifies to
\begin{equation}
\frac{\partial \mathcal{L}_{\text{aux}}}{\partial z_j(v)}
= P_j(v) - \tilde{y}_j(v).
\end{equation}
The derivation process is provided in the Appendix \ref{sec:aux_grad}.

\textbf{Gradient of soft NTP.}
In contrast, for Eq.~\ref{eq:soft_label} corresponds to
\begin{equation}
\mathcal{L}_{\text{soft-NTP}}
= - \sum_{j=1}^{|t|} \log \sum_{v\in\mathcal V}\tilde{y}_j(v)\,P_j(v).
\end{equation}
Its gradient is given by
\begin{equation}
\begin{aligned}
\nabla_\theta \mathcal{L}_{\text{soft-NTP}}
&= - \sum_{j=1}^{|t|}\sum_{v\in\mathcal V} q_j(v)\,\nabla_\theta \log P_j(v),
\\
q_j(v) &= \frac{\tilde{y}_j(v)\,P_j(v)}{\sum_{u}\tilde{y}_j(u)\,P_j(u)}.
\end{aligned}
\end{equation}
Equivalently, in logit space we have
\begin{equation}
\frac{\partial \mathcal{L}_{\text{soft-NTP}}}{\partial z_j(v)}
= P_j(v) - q_j(v).
\end{equation}
For details of the derivation, please refer to Appendix \ref{sec:softNTP_grad}.

\textbf{Discussion.}
Although the auxiliary KL and soft-NTP objectives appear similar, their gradient weighting schemes differ fundamentally. 
In the auxiliary KL formulation, the gradient is driven by the fixed target distribution $\tilde{y}_j$, where model updates would move toward the provided supervisory signal. 
In contrast, the soft-NTP formulation replaces the fixed weights with the adaptive distribution $q_j$, which depends jointly on $\tilde{y}_j$ and the model’s current prediction $P_j$. 
This adaptivity shifts provide a more comprehensive and balanced consideration of collaborative signals and the world knowledge from LLMs. 



\section{Experiment}

\subsection{Settings}
\subsubsection{Dataset}
 To validate the effectiveness of our approach, we used three widely adopted publicly available datasets: Toys, Sports and Office. The statistics of these dataset is listed in Table \ref{tb:statistics}. All these three datasets can be found on the website\footnote{https://cseweb.ucsd.edu/~jmcauley/datasets/amazon\_v2/}. 
\begin{itemize}
    \item \textbf{Toys} encompasses user reviews for toys on Amazon, including reviewer ID, item ID, title, and brands, etc.
    \item \textbf{Sports} contains user feedback and ratings for a wide variety of products in sports.  
    \item \textbf{Office} includes reviews and scores for Office, reflecting how users rate different office-related items online.  
\end{itemize}
Due to the resource-intensive nature of LLMs, it is crucial to balance computational cost with performance. Thus, we apply 5-core settings across three datasets, that is, we filter out the user and items that is less than 5 interaction. In addition, we ensure that each interaction is associated with at least 10 historical items, so as to observe the performance of different models under sufficient contextual information. We then split the data in an 8:1:1 ratio into training, validation, and test sets, maintaining a chronological order to prevent data leakage. This ensures that the reported performance is both accurate and reliable.
More detailed statistical information of three datasets is provided in Table. \ref{tb:statistics}.

\begin{table}
\centering
\caption{Statistics of Datasets.}
\label{tb:statistics}
\begin{tabular}{cccccc}
\hline
Dataset & \#User & \#Item & \#Interactions & \#Density \\ \hline
Toys           & 19,513              &  20,669         & 117,131 &  0.029\%       \\ 
Sports      & 29,109                & 27,008         & 161,564  &  0.025\%      \\
Office      & 12,603               & 9,913         & 70,869    &  0.056\%   \\

\hline
\end{tabular}

\end{table} 
\begin{table*}[t]
\caption{The experimental comparison among a wide range of recommendation approaches for three datasets. N@k is the shorthand of NDCG, and H@k means Hit Ratio. "+TCA" indicates that the corresponding baseline is enhanced with our proposed method. The best results are in bold and the secondary best results are underlined.}
\label{tab:overall_performance}
\centering
\setlength{\tabcolsep}{2.0mm}{
\begin{tabular}{p{1.8cm}|cccc|cccc|cccc}
\hline
       Model     & \multicolumn{4}{c|}{Toys}              &  \multicolumn{4}{c|}{Sports}   &  \multicolumn{4}{c}{Office}       

\\
\hline
            & N@5 & N@10 & H@5 & H@10              &  N@5 & N@10 & H@5 & H@10   &  N@5 & N@10 & H@5 & H@10       
        
\\
\hline 
LightGCN
 & 0.0005 & 0.0009 & 0.0017 & 0.0027
 & 0.0006 & 0.0011 & 0.0011 & 0.0028
 & 0.0036 & 0.0045 & 0.0076 & 0.0119
\\
 SASRec  
 & 0.0218 & 0.0247 & \underline{0.0316} & \underline{0.0406} 
 & 0.0207 & 0.0232 & 0.0305 & 0.0378	
 & 0.0497 & 0.0524 & \textbf{0.0708} & \underline{0.0792}	
\\
 BERT4Rec  
 & 0.0132 & 0.0174 & 0.0188 & 0.0304 
 & 0.0112 & 0.0142 & 0.0170 & 0.0283 
 & 0.0287 & 0.0327 & 0.0392 & 0.0546
\\
\hline
 AlphaRec  
 & 0.0010 & 0.0013 & 0.0022 & 0.0041
 & 0.0013 & 0.0031 & 0.0025 & 0.0083
 & 0.0024 & 0.0071 & 0.0088 & 0.0111
\\
 RLMRec  
 & 0.0054 & 0.0076 & 0.0090 & 0.0160	& 0.0049 & 0.0065 & 0.0077 & 0.0128	& 0.0104 & 0.0131 & 0.0147 & 0.0231
\\
\hline
 TallRec  
 & 0.0099 & 0.0116 & 0.0120 & 0.0170 
 & 0.0085 & 0.0097 & 0.0105 & 0.0142
 & 0.0380 & 0.0427 & 0.0494 & 0.0640
 \\
 + TCA  
 & 0.0224 & \underline{0.0249} & 0.0292 & 0.0368	
 & \underline{0.0281} & \underline{0.0299} & \underline{0.0324} & \underline{0.0379}	
 & 0.0451 & 0.0475 & 0.0552 & 0.0626
\\
\hline
 Llara  
 & 0.0161 & 0.0182 & 0.0212 & 0.0276
 & 0.0186 & 0.0208 & 0.0236 & 0.0306
 & 0.0383 & 0.0424 & 0.0512 & 0.0638
\\
 + TCA  
 & \underline{0.0228} & 0.0244 & 0.0302 & 0.0350
 & 0.0262 & 0.0286 & 0.0304 & 0.0376
 & 0.0490 & 0.0529 & 0.0614 & 0.0736
 \\
\hline
 Collm  
 & 0.0160 & 0.0186 & 0.0209 & 0.0292
 & 0.0084 & 0.0101 & 0.0108 & 0.0162
 & 0.0398 & 0.0437 & 0.0524 & 0.0641
\\
 + TCA  
 & 0.0206 & 0.0227 & 0.0260 & 0.0324
 & 0.0260 & 0.0275 & 0.0292 & 0.0335	
 & 0.0476 & 0.0520 & 0.0604 & 0.0744
 \\
\hline
 MSL  
 & 0.0145 & 0.0175 & 0.0204 & 0.0297
 & 0.0259 & 0.0272 & 0.0314 & 0.0356
 & \underline{0.0509} & \underline{0.0547} & 0.0662 & 0.0780
\\
 + TCA  
 & \textbf{0.0332} & \textbf{0.0375} & \textbf{0.0452} & \textbf{0.0584}
 & \textbf{0.0309} & \textbf{0.0338} & \textbf{0.0370} & \textbf{0.0456}	
 & \textbf{0.0544} & \textbf{0.0584} & \underline{0.0692} & \textbf{0.0813}
 \\
\hline
\end{tabular}
}

\end{table*}
\subsubsection{Baselines}
To conduct a more comprehensive analysis of the capabilities of our method, we introduce various baselines to provide fair comparisons and highlight the relative strengths of our approach. In addition, since our method can be seamlessly integrated with any LLM-based generative recommendation approach, we further enhance four representative LLM-based recommendation algorithms. The detailed introduction of these baselines refers to Appendix \ref{sec:baseline}.

Due to space limitations, detailed implementation settings and evaluation metrics refer to Appendix \ref{sec:exp_setting}.


\subsection{Overall Performance}
In this section, we conducted numerous experiments on three publicly available datasets and demonstrated the superior performance of our method. The detailed results are reported in Table \ref{tab:overall_performance}. From that, we can conclude the following observations:
\begin{itemize}
    \item Our method surpasses all strong and state-of-the-art baselines by a large margin. Furthermore, it consistently outperforms the vanilla base models, including TallRec, Llara, Collm, MSL, showing its universality and applicability across diverse backbone architectures and effectiveness in enhancing the recommendation capability of LLMs.
    \item The series of models that introduce adapters to inject CF signals into the LLM’s soft prompts do not always yield improvements. Specifically, on the Office dataset, CollM and LLaRA show little advantage over TallRec after incorporating CF signals, and on the Sports dataset, CollM even performs slightly worse than TallRec.
    \item Non-sequential models show  incompatibility with our task, even AlphaRec and RLMRec, which leverages LLM embeddings of user profile and title to do the recommendation task. This can be attributed to the cold-start nature of our setting, where the test set contains many unseen users and items, making it challenging for non-sequential models to generalize effectively. 
    \item Traditional sequential recommendation methods like SASRec still remain highly effective, outperforming all LLM-based generative approaches except ours. This indicates that current LLM-based generative methods still struggle to fully capture recommendation capability, further underscoring the necessity of exploring how to incorporate collaborative filtering signals.
\end{itemize}

\begin{figure*}[htbp]
\centering
\caption{Relation between collaborative consistency and performance. The x-axis refers to the hyperparameter $\alpha$. The dashed lines denote the collaborative consistency of three models.}
\label{fig.rics}
\includegraphics[width=\textwidth ]{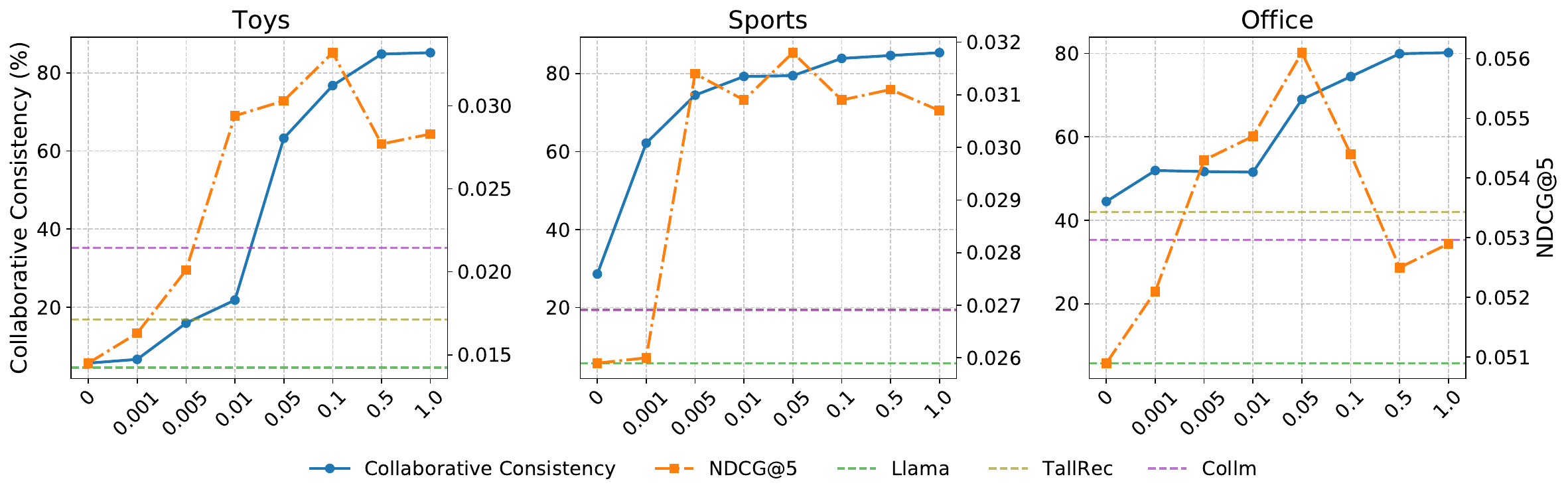}
\end{figure*}
\subsection{Model-Agnostic Capability}
To validate the model-agnostic capability of our approach, in this subsection, we evaluate the performance of TCA when applied to quantization models. Specifically, we adopt representative quantization models, namely TIGER \cite{rajput2023recommender} and LETTER \cite{wang2024learnable}. Following the experimental setup of TIGER, we use an encoder-decoder framework called sentence-T5 \cite{ni2022sentence} as the LLM backbone. To ensure fairness, we only use item title to generate item semantic embedding via Llama3.1-3b. Other experimental settings are consistent with the our previous settings.

As illustrated in Table 3, we draw the following conclusions.
(1) TCA consistently improves the performance of SID–based methods across all three datasets. This indicates that our approach can be effectively applied to different LLM backbones, demonstrating strong model-agnostic capability.
(2) LETTER consistently outperforms TIGER. This is because LETTER incorporates collaborative filtering signals during Semantic ID generation, resulting in higher-quality Semantic IDs and consequently better recommendation performance.
(3) Overall, SID-based methods exhibit relatively limited performance. Since only item titles are available for Semantic ID generation in our setting, the quality of the generated IDs is limited, which constrains the upper bound of recommendation performance.
\begin{table}
\hspace{3mm}
\caption{Performance comparison of TIGER, LETTER, and TCA-enhanced models across three datasets. }
\label{tab:sid}
\setlength{\tabcolsep}{1.0mm}{
\begin{tabular}{p{1.5cm}|cc|cc|cc}
\hline
       \multirow{2}{*}{Model}     & \multicolumn{2}{c|}{Toys}              &  \multicolumn{2}{c|}{Sports} &
       \multicolumn{2}{c}{Office}
\\
 & N@5 & H@5 & N@5 & H@5 & N@5 & H@5
 \\
\hline
SASRec  
 & 0.0218  & 0.0316 
 & 0.0207  & 0.0305 
 & 0.0497  & 0.0708  \\
\hline
\hline

TIGER
&  0.0031 & 0.0040
 & 0.0027  & 0.0042
 & 0.0176  & 0.0242
\\
\hline
+TCA
& 0.0045  & 0.0068
 & 0.0040  & 0.0068
 & 0.0176  & 0.0246
\\
\hline
\hline
LETTER
&  0.0052  & 0.0076
 & 0.0031  & 0.0056
 & 0.0170 & 0.0248 
\\
\hline
+TCA
& \textbf{0.0068}  & \textbf{0.0098} 
 & \textbf{0.0049}  & \textbf{0.0086}
 & \textbf{0.0246}  & \textbf{0.0310}
\\
\hline
\end{tabular}
}

\end{table}
\subsection{Collaborative Consistency}
Beyond superior performance, it is also important to verify whether the observed information gain stems from better alignment with collaborative filtering signals. To assess this, we introduce a new metric, termed collaborative consistency (CC), which measures the extent to which our predictions remain consistent with the traditional recommender model. Specifically, we compare whether the top-1 prediction of a traditional CF model matches the top-1 prediction of our LLM-based approach. A higher collaborative consistency score indicates that our method is more inclined to align with the predictions of the traditional model. This experiment is conducted based on the MSL \cite{wang2025msltokensneedtuning} model and the results are listed in Figure~\ref{fig.rics}. Note that the x-axis refers to the weighting hyperparameter $\alpha$. From the figure, some insight can be observed:
\begin{itemize}
    \item Collaborative consistency increases steadily as $\alpha$ grows, suggesting that a larger weight on the collaborative distribution leads the model to generate predictions more aligned with traditional CF signals. This demonstrates that our proposed method can effectively enable the LLM to perceive and incorporate collaborative filtering signals, which is well aligned with our motivation.
    \item  The performance of models increases rapidly as collaborative consistency improves, but after reaching a peak, further enlarging $\alpha$ deteriorates overall performance. It is intuitive that while CF signals are crucial for recommendation, they are not always reliable and often contain noise and errors. In this line, a small $\alpha$ allows the LLM to explore diverse possibilities, whereas an excessively large $\alpha$ introduces more noise from CF, ultimately degrading performance.
    \item The vanilla LLaMA3 model yields low scores in collaborative consistency, suggesting that LLMs initially lack an understanding of recommendation tasks. After fine-tuning (TallRec), CC scores rise substantially even without collaborative signals, e.g., 19.54\% on Sports and 42.01\% on Office, indicating that fine-tuning itself effectively adapts the model to recommendation and thereby improves performance.
    
\end{itemize} 
\begin{table}[t]
\end{table}

\subsection{Ablation Study}
Table \ref{tab:ablation_study} presents the results of our ablation study, where we investigate the  effects of the Collaborative Tokenizer (CT) and Soft-Label Alignment (SA) modules, respectively. 

\textbf{w/o CT variant}. Since the Collaborative Tokenizer transforms item-level collaborative information into a token-level probability distribution, removing CT degenerates this distribution into a uniform one, thereby eliminating item-specific collaborative signals at the token level.

\textbf{w/o SA variant}. Removing the Soft-Label Alignment (SA) module prevents TCA from integrating CF information into the loss function. As described in Section \ref{sec:kl_loss}, the auxiliary KL loss is closely related to soft-label supervision. Accordingly, in the w/o SA variant, we adopt the \textbf{auxiliary KL loss} in place of SA.

The results reveal several key insights:
(1) The CT plays a critical role in our framework. Removing CT leads to a substantial performance drop, as the resulting uniform token-level distribution degenerates into pure noise and fails to provide informative signals to the LLM.
(2) The auxiliary KL loss remains effective to some extent. As shown in the results, the variant using the KL loss still outperforms the MSL baseline, indicating that the KL objective can absorb and exploit useful collaborative information.
(3) SA performs better overall. Although the KL loss contributes positively, the soft-label technique consistently achieves superior performance, which empirically validates our prior theoretical analysis.
\begin{table}
\hspace{3mm}
\caption{Ablation Study. \textbf{w/o CT} represents removing Collaborative Tokenizer module and \textbf{w/o SA} denotes removing the Soft-Label Alignment module. }
\label{tab:ablation_study}
\setlength{\tabcolsep}{1.0mm}{
\begin{tabular}{p{1.5cm}|cc|cc|cc}
\hline
       \multirow{2}{*}{Model}     & \multicolumn{2}{c|}{Toys}              &  \multicolumn{2}{c|}{Sports} &
       \multicolumn{2}{c}{Office}
\\
 & N@5 & H@5 & N@5 & H@5 & N@5 & H@5
 \\
\hline
MSL  
 & 0.0145  & 0.0204 
 & 0.0259  & 0.0314 
 & 0.0509  & 0.0662  \\
\hline
\hline

w/o CT
&  0.0181 & 0.0238
 & 0.0189  & 0.0221
 & 0.0424  & 0.0557
\\
\hline
w/o SA
& 0.0276  & 0.0381 
 & 0.0301  & 0.0351
 & 0.0509  & 0.0653
\\
\hline

TCA
& \textbf{0.0332}  & \textbf{0.0452} 
 & \textbf{0.0309}  & \textbf{0.0370}
 & \textbf{0.0544}  & \textbf{0.0692}
\\
\hline
\end{tabular}
}

\end{table}
\subsection{Efficiency of Incorporating CF Signals}
As previously stated, our key motivation is to identify an effective way to introduce CF signals into LLMs. To verify whether our method possesses this capability, we examine how model performance varies with the strength of the incorporated CF signals. 
Specifically, there are two key factors that influence the quality of CF signals: the size of training data and the choice of different backbones.

\subsubsection{Training Size} 
In this part, we investigate how the training size of CF models affects the final performance. Specifically, we train SASRec on datasets of different sizes (10K, 50K, and the full dataset) using the Toys dataset.
We further include Collm as a representative baseline for CF signal incorporation. Notably, only the training size of SASRec is varied, while Collm and TCA4Rec are trained on a fixed dataset with 10K samples. 

As shown in Table \ref{tab:efficiency}, we make the following observations. (1) With increasing training data, the performance of SASRec steadily improves, which in turn results in stronger collaborative filtering signals. (2) As the CF signals become stronger, the performance of TCA4Rec improves accordingly, demonstrating its ability to efficiently absorb collaborative knowledge and thereby validating our core motivation. (3) In contrast, despite the continuous strengthening of CF signals, Collm fails to benefit from them and even shows a slight performance degradation, indicating its limited effectiveness in leveraging collaborative filtering signals.
\subsubsection{CF Models}
Different CF models produce fundamentally different CF signals. Therefore, it is crucial to explore how the choice of CF models affects overall performance. To this end, we adopt another popular CF model BERT4Rec to conduct the experiment under the MSL framework. 

As shown in Table \ref{tab:backbone}, we observe that (1) Stronger CF models lead to greater performance improvements. Consistent with the observations on training size, higher-quality CF signals more effectively enhance the performance of TCA. (2) Weaker CF models may impair the effectiveness of TCA. When quality of CF model degrades, the resulting collaborative signals become less reliable. The models that absorb such noisy or misleading information would in turn reduces their overall performance.
\begin{table}
\hspace{3mm}
\caption{Effect of CF model training size on LLM4Rec performance. Small, Medium, Large denotes the different training size. Note that only SASRec is trained with varying data sizes, while Collm and TCA4Rec are still trained on the fixed set of 10K training samples.}
\label{tab:efficiency}
\setlength{\tabcolsep}{1.0mm}{
\begin{tabular}{p{1.5cm}|cc|cc|cc}
\hline
       \multirow{2}{*}{Model}     & \multicolumn{2}{c|}{Small Data}              &  \multicolumn{2}{c|}{Medium Data} &
       \multicolumn{2}{c}{Large Data}
\\
 & N@5 & H@5 & N@5 & H@5 & N@5 & H@5
 \\
\hline
SASRec  
 & 0.0126  & 0.0130 
 & 0.0196  & 0.0274 
 & 0.0218  & 0.0316  \\
\hline
\hline
+Collm
& 0.0180  & 0.0246
 & 0.0160  & 0.0209
 & 0.0161  & 0.0212
\\
\hline

+TCA
& \textbf{0.0234}  & \textbf{0.0324} 
 & \textbf{0.0309}  & \textbf{0.0370}
 & \textbf{0.0332}  & \textbf{0.0452}
\\
\hline
\end{tabular}
}

\end{table}


\begin{table}
\hspace{3mm}
\caption{Impact of different backbones.}
\label{tab:backbone}
\setlength{\tabcolsep}{1.0mm}{
\begin{tabular}{p{1.5cm}|cc|cc|cc}
\hline
       \multirow{2}{*}{Model}     & \multicolumn{2}{c|}{Toys}              &  \multicolumn{2}{c|}{Sports} &
       \multicolumn{2}{c}{Office}
\\
 & N@5 & H@5 & N@5 & H@5 & N@5 & H@5
 \\
\hline
MSL  
 & 0.0145  & 0.0204 
 & 0.0259  & 0.0314 
 & 0.0509  & 0.0662  \\
\hline
\hline
BERT4Rec  
 & 0.0132 & 0.0188 & 0.0112  & 0.0170 & 0.0287  & 0.0392 \\
 +TCA
 & 0.0222 & 0.0297 & 0.0127 &	0.0153 & 0.0347 & 0.0443 \\
\hline
SASRec
& 0.0218 &  0.0316 & 0.0207 & 0.0305
 & 0.0497 &  \textbf{0.0708} 	
\\
+TCA
& \textbf{0.0332}  & \textbf{0.0452} 
 & \textbf{0.0309}  & \textbf{0.0370}
 & \textbf{0.0544}  & 0.0692
\\
\hline
\end{tabular}
}

\end{table}

\section{Related Work}
\subsection{LLM-based Recommendation}
Over the past two years, LLM-based recommendation methods have sprung out rapidly as its impressive reasoning capability drives researchers to exert considerable effort in leveraging LLMs for recommendation systems \cite{zhu2024recommender,DBLP:journals/www/WuZQWGSQZZLXC24,DBLP:journals/corr/abs-2410-19744,zheng2024harnessing,lin2025recommendation}. Generally, existing methods can be divided into two categories.
\subsubsection{LLM as Recommender} The goal of this line is to use generative tasks to produce recommendation results \cite{DBLP:journals/corr/abs-2305-00447,kong2024customizing}. Initially, researchers applied In-Context Learning (ICL) to equip large models with recommendation capabilities. Later, approaches like TallRec \cite{DBLP:journals/corr/abs-2305-00447} adopted LoRA \cite{DBLP:conf/iclr/HuSWALWWC22} for fine-tuning, aiming to align the recommendation systems with LLMs. 
As the prevailing of MoE \cite{DBLP:journals/corr/abs-2408-10159} technique, iLoRA  \cite{kong2024customizing} decomposes LoRA into multiple smaller instances to represent user multifaceted interests. Then, MSL  \cite{wang2025msltokensneedtuning} masks invalid tokens during loss computation to focus on beneficial tokens. More recently, HatLLM \cite{cui2025hatllm} further introduce hierarchical attention masking to enable better modeling of both semantic and collaborative signals in sequential data.
Although these methods are effective, they still struggle to explicitly capture collaborative signals between users and items.

\subsubsection{LLM as enhancer.}
In this line, the LLMs serves as a multimodal encoder to obtain more fine-grained representations that allow downstream tasks to access a broader scope of knowledge \cite{wang2025llm4dsr,cui2024distillation,lin2024box}. The emergence of BERT \cite{DBLP:conf/naacl/DevlinCLT19} revealed the remarkable text comprehension capabilities of general large models. 
These findings sparked the development of several alignment frameworks, such as contrastive learning \cite{DBLP:conf/mm/WeiWLNLLC21}, which target at integrating these high-quality embeddings into existing recommender systems.
With the advent of LLMs, more researchers had sought to bring their understanding capability into recommendation systems, like LLMEmb \cite{DBLP:journals/corr/abs-2409-19925} and AlphaRec \cite{DBLP:journals/corr/abs-2407-05441}. 
Although promising, they overlook the powerful reasoning ability of LLMs, which is considered as the key to propel LLMs to global prominence. 


\subsection{Collaborative Signals in LLMRec}
Although LLMs encode rich world knowledge, they still struggle to learn collaborative knowledge from interaction data due to its inherent sparsity and noise. To address this, a promising approach is to incorporate collaborative signals from traditional recommendation models, which has inspired a series of effective hybrid methods.
For example, Collm \cite{DBLP:journals/corr/abs-2310-19488} is one of the earliest work, where collaborative signals are considered as continuous prompt embeddings into LLMs embeddings. LLaRA  \cite{liao2024llara} extends this work and introduces curriculum learning to better adapt the model to the newly introduced tokens. BinLLM \cite{zhang2024text} tries to transforms collaborative embeddings into compact binary codes and further maps them into text-like sequences, allowing LLMs to incorporate collaborative signals through pure textual inputs.
Recently, a new fusion paradigm has attracted increasing attention. TIGER \cite{rajput2023recommender} introduces a novel tokenizer module, where item embeddings are decomposed into semantic IDs via RQ-VAE and then fed into LLMs for training. Building on this idea, LC-Rec \cite{zheng2024adapting} incorporates additional alignment objectives to help LLMs better understand semantic IDs. LETTER \cite{wang2024learnable} further integrates collaborative filtering signals into the tokenizer module, enabling tighter coupling between tokenization and collaborative information. ETEGRec \cite{liu2025generative} unifies item tokenization and recommendation training in a joint end-to-end manner, allowing the tokenizer to adapt to user interaction patterns dynamically.
However, these methods still exhibit limitations in explicitly and sufficiently modeling collaborative signals in recommendation tasks.
\section{Conclusion}
In this paper, we proposed a model-agnostic framework, Token-level Collaborative Alignment for Recommendation (TCA4Rec) that injects collaborative filtering signals into LLMs, including Collaborative Tokenizer and Soft Label Alignment modules. This design bridges the gap between item-level CF preferences and token-level supervision, allowing LLMs to capture user–item relations while preserving generative ability. Experiments show consistent improvements over strong baselines, and further validate that our method enhances alignment with CF signals.

\begin{acks}
This work was supported in part by the grants from National Science and Technology Major Project (No. 2023ZD0121104), National Natural Science Foundation of China (No.62222213, U22B2059), in part by the Postdoctoral Fellowship Program and China Postdoctoral Science Foundation under Grant Number BX20250387 and 2025M781529, in part by the Fundamental Research Funds for the Central Universities under Grant Number WK2150250042.
\end{acks}

\bibliographystyle{ACM-Reference-Format}
\bibliography{sample-base}

@String{Computer = "{IEEE} Computer" }

@String{Springer = "Springer-Verlag" }

@inproceedings{DBLP:conf/sigir/0001DWLZ020,
  author       = {Xiangnan He and
                  Kuan Deng and
                  Xiang Wang and
                  Yan Li and
                  Yong{-}Dong Zhang and
                  Meng Wang},
  editor       = {Jimmy X. Huang and
                  Yi Chang and
                  Xueqi Cheng and
                  Jaap Kamps and
                  Vanessa Murdock and
                  Ji{-}Rong Wen and
                  Yiqun Liu},
  title        = {LightGCN: Simplifying and Powering Graph Convolution Network for Recommendation},
  booktitle    = {Proceedings of the 43rd International {ACM} {SIGIR} conference on
                  research and development in Information Retrieval, {SIGIR} 2020, Virtual
                  Event, China, July 25-30, 2020},
  pages        = {639--648},
  publisher    = {{ACM}},
  year         = {2020},
  url          = {https://doi.org/10.1145/3397271.3401063},
  doi          = {10.1145/3397271.3401063},
  bibsource    = {dblp computer science bibliography, https://dblp.org}
}

@article{DBLP:journals/corr/abs-2310-19488,
  author       = {Yang Zhang and
                  Fuli Feng and
                  Jizhi Zhang and
                  Keqin Bao and
                  Qifan Wang and
                  Xiangnan He},
  title        = {CoLLM: Integrating Collaborative Embeddings into Large Language Models
                  for Recommendation},
  journal      = {CoRR},
  volume       = {abs/2310.19488},
  year         = {2023},
  url          = {https://doi.org/10.48550/arXiv.2310.19488},
  doi          = {10.48550/ARXIV.2310.19488},
  eprinttype    = {arXiv},
  eprint       = {2310.19488},
  bibsource    = {dblp computer science bibliography, https://dblp.org}
}

@article{zhu2024recommender,
  title={Recommender Systems Meet Large Language Model Agents: A Survey},
  auCthor={Zhu, Xi and Wang, Yu and Gao, Hang and Xu, Wujiang and Wang, Chen and Liu, Zhiwei and Wang, Kun and Jin, Mingyu and Pang, Linsey and Wen, Qingsong and others},
  journal={Available at SSRN 5062105},
  year={2024}
}

@article{DBLP:journals/www/WuZQWGSQZZLXC24,
  author       = {Likang Wu and
                  Zhi Zheng and
                  Zhaopeng Qiu and
                  Hao Wang and
                  Hongchao Gu and
                  Tingjia Shen and
                  Chuan Qin and
                  Chen Zhu and
                  Hengshu Zhu and
                  Qi Liu and
                  Hui Xiong and
                  Enhong Chen},
  title        = {A survey on large language models for recommendation},
  journal      = {World Wide Web {(WWW)}},
  volume       = {27},
  number       = {5},
  pages        = {60},
  year         = {2024},
  url          = {https://doi.org/10.1007/s11280-024-01291-2},
  doi          = {10.1007/S11280-024-01291-2},
  bibsource    = {dblp computer science bibliography, https://dblp.org}
}

@article{DBLP:journals/corr/abs-2410-19744,
  author       = {Qi Wang and
                  Jindong Li and
                  Shiqi Wang and
                  Qianli Xing and
                  Runliang Niu and
                  He Kong and
                  Rui Li and
                  Guodong Long and
                  Yi Chang and
                  Chengqi Zhang},
  title        = {Towards Next-Generation LLM-based Recommender Systems: {A} Survey
                  and Beyond},
  journal      = {CoRR},
  volume       = {abs/2410.19744},
  year         = {2024},
  url          = {https://doi.org/10.48550/arXiv.2410.19744},
  doi          = {10.48550/ARXIV.2410.19744},
  eprinttype    = {arXiv},
  eprint       = {2410.19744},
  bibsource    = {dblp computer science bibliography, https://dblp.org}
}

@article{DBLP:journals/corr/abs-2305-00447,
  author       = {Keqin Bao and
                  Jizhi Zhang and
                  Yang Zhang and
                  Wenjie Wang and
                  Fuli Feng and
                  Xiangnan He},
  title        = {TALLRec: An Effective and Efficient Tuning Framework to Align Large
                  Language Model with Recommendation},
  journal      = {CoRR},
  volume       = {abs/2305.00447},
  year         = {2023},
  url          = {https://doi.org/10.48550/arXiv.2305.00447},
  doi          = {10.48550/ARXIV.2305.00447},
  eprinttype    = {arXiv},
  eprint       = {2305.00447},
  bibsource    = {dblp computer science bibliography, https://dblp.org}
}

@article{kong2024customizing,
  title={Customizing language models with instance-wise lora for sequential recommendation},
  author={Kong, Xiaoyu and Wu, Jiancan and Zhang, An and Sheng, Leheng and Lin, Hui and Wang, Xiang and He, Xiangnan},
  journal={arXiv preprint arXiv:2408.10159},
  year={2024}
}

@inproceedings{liao2024llara,
  title={Llara: Large language-recommendation assistant},
  author={Liao, Jiayi and Li, Sihang and Yang, Zhengyi and Wu, Jiancan and Yuan, Yancheng and Wang, Xiang and He, Xiangnan},
  booktitle={Proceedings of the 47th International ACM SIGIR Conference on Research and Development in Information Retrieval},
  pages={1785--1795},
  year={2024}
}

@inproceedings{DBLP:conf/iclr/HuSWALWWC22,
  author       = {Edward J. Hu and
                  Yelong Shen and
                  Phillip Wallis and
                  Zeyuan Allen{-}Zhu and
                  Yuanzhi Li and
                  Shean Wang and
                  Lu Wang and
                  Weizhu Chen},
  title        = {LoRA: Low-Rank Adaptation of Large Language Models},
  booktitle    = {The Tenth International Conference on Learning Representations, {ICLR}
                  2022, Virtual Event, April 25-29, 2022},
  publisher    = {OpenReview.net},
  year         = {2022},
  url          = {https://openreview.net/forum?id=nZeVKeeFYf9},
  bibsource    = {dblp computer science bibliography, https://dblp.org}
}

@article{DBLP:journals/corr/abs-2408-10159,
  author       = {Xiaoyu Kong and
                  Jiancan Wu and
                  An Zhang and
                  Leheng Sheng and
                  Hui Lin and
                  Xiang Wang and
                  Xiangnan He},
  title        = {Customizing Language Models with Instance-wise LoRA for Sequential
                  Recommendation},
  journal      = {CoRR},
  volume       = {abs/2408.10159},
  year         = {2024},
  url          = {https://doi.org/10.48550/arXiv.2408.10159},
  doi          = {10.48550/ARXIV.2408.10159},
  eprinttype    = {arXiv},
  eprint       = {2408.10159},
  bibsource    = {dblp computer science bibliography, https://dblp.org}
}

@article{DBLP:journals/corr/abs-2406-10450,
  author       = {Haohao Qu and
                  Wenqi Fan and
                  Zihuai Zhao and
                  Qing Li},
  title        = {TokenRec: Learning to Tokenize {ID} for LLM-based Generative Recommendation},
  journal      = {CoRR},
  volume       = {abs/2406.10450},
  year         = {2024},
  url          = {https://doi.org/10.48550/arXiv.2406.10450},
  doi          = {10.48550/ARXIV.2406.10450},
  eprinttype    = {arXiv},
  eprint       = {2406.10450},
  bibsource    = {dblp computer science bibliography, https://dblp.org}
}

@inproceedings{DBLP:conf/naacl/DevlinCLT19,
  author       = {Jacob Devlin and
                  Ming{-}Wei Chang and
                  Kenton Lee and
                  Kristina Toutanova},
  editor       = {Jill Burstein and
                  Christy Doran and
                  Thamar Solorio},
  title        = {{BERT:} Pre-training of Deep Bidirectional Transformers for Language
                  Understanding},
  booktitle    = {Proceedings of the 2019 Conference of the North American Chapter of
                  the Association for Computational Linguistics: Human Language Technologies,
                  {NAACL-HLT} 2019, Minneapolis, MN, USA, June 2-7, 2019, Volume 1 (Long
                  and Short Papers)},
  pages        = {4171--4186},
  publisher    = {Association for Computational Linguistics},
  year         = {2019},
  url          = {https://doi.org/10.18653/v1/n19-1423},
  doi          = {10.18653/V1/N19-1423},
  bibsource    = {dblp computer science bibliography, https://dblp.org}
}

@inproceedings{DBLP:conf/mm/WeiWLNLLC21,
  author       = {Yinwei Wei and
                  Xiang Wang and
                  Qi Li and
                  Liqiang Nie and
                  Yan Li and
                  Xuanping Li and
                  Tat{-}Seng Chua},
  editor       = {Heng Tao Shen and
                  Yueting Zhuang and
                  John R. Smith and
                  Yang Yang and
                  Pablo C{\'{e}}sar and
                  Florian Metze and
                  Balakrishnan Prabhakaran},
  title        = {Contrastive Learning for Cold-Start Recommendation},
  booktitle    = {{MM} '21: {ACM} Multimedia Conference, Virtual Event, China, October
                  20 - 24, 2021},
  pages        = {5382--5390},
  publisher    = {{ACM}},
  year         = {2021},
  url          = {https://doi.org/10.1145/3474085.3475665},
  doi          = {10.1145/3474085.3475665},
  bibsource    = {dblp computer science bibliography, https://dblp.org}
}

@inproceedings{zhang2024notellm,
  title={NoteLLM: A Retrievable Large Language Model for Note Recommendation},
  author={Zhang, Chao and Wu, Shiwei and Zhang, Haoxin and Xu, Tong and Gao, Yan and Hu, Yao and Chen, Enhong},
  booktitle={Companion Proceedings of the ACM on Web Conference 2024},
  pages={170--179},
  year={2024}
}

@article{DBLP:journals/corr/abs-2407-05441,
  author       = {Leheng Sheng and
                  An Zhang and
                  Yi Zhang and
                  Yuxin Chen and
                  Xiang Wang and
                  Tat{-}Seng Chua},
  title        = {Language Models Encode Collaborative Signals in Recommendation},
  journal      = {CoRR},
  volume       = {abs/2407.05441},
  year         = {2024},
  url          = {https://doi.org/10.48550/arXiv.2407.05441},
  doi          = {10.48550/ARXIV.2407.05441},
  eprinttype    = {arXiv},
  eprint       = {2407.05441},
  bibsource    = {dblp computer science bibliography, https://dblp.org}
}

@article{DBLP:journals/corr/abs-2409-19925,
  author       = {Qidong Liu and
                  Xian Wu and
                  Wanyu Wang and
                  Yejing Wang and
                  Yuanshao Zhu and
                  Xiangyu Zhao and
                  Feng Tian and
                  Yefeng Zheng},
  title        = {Large Language Model Empowered Embedding Generator for Sequential
                  Recommendation},
  journal      = {CoRR},
  volume       = {abs/2409.19925},
  year         = {2024},
  url          = {https://doi.org/10.48550/arXiv.2409.19925},
  doi          = {10.48550/ARXIV.2409.19925},
  eprinttype    = {arXiv},
  eprint       = {2409.19925},
  bibsource    = {dblp computer science bibliography, https://dblp.org}
}

@inproceedings{DBLP:conf/icdm/KangM18,
  author       = {Wang{-}Cheng Kang and
                  Julian J. McAuley},
  title        = {Self-Attentive Sequential Recommendation},
  booktitle    = {{IEEE} International Conference on Data Mining, {ICDM} 2018, Singapore,
                  November 17-20, 2018},
  pages        = {197--206},
  publisher    = {{IEEE} Computer Society},
  year         = {2018},
  url          = {https://doi.org/10.1109/ICDM.2018.00035},
  doi          = {10.1109/ICDM.2018.00035},
  bibsource    = {dblp computer science bibliography, https://dblp.org}
}

@article{sheng2024language,
  title={Language Representations Can be What Recommenders Need: Findings and Potentials},
  author={Sheng, Leheng and Zhang, An and Zhang, Yi and Chen, Yuxin and Wang, Xiang and Chua, Tat-Seng},
  journal={arXiv preprint arXiv:2407.05441},
  year={2024}
}

@inproceedings{sun2019bert4rec,
  title={BERT4Rec: Sequential recommendation with bidirectional encoder representations from transformer},
  author={Sun, Fei and Liu, Jun and Wu, Jian and Pei, Changhua and Lin, Xiao and Ou, Wenwu and Jiang, Peng},
  booktitle={Proceedings of the 28th ACM international conference on information and knowledge management},
  pages={1441--1450},
  year={2019}
}

@inproceedings{ren2024representation,
  title={Representation learning with large language models for recommendation},
  author={Ren, Xubin and Wei, Wei and Xia, Lianghao and Su, Lixin and Cheng, Suqi and Wang, Junfeng and Yin, Dawei and Huang, Chao},
  booktitle={Proceedings of the ACM web conference 2024},
  pages={3464--3475},
  year={2024}
}

@article{dubey2024llama,
  title={The llama 3 herd of models},
  author={Dubey, Abhimanyu and Jauhri, Abhinav and Pandey, Abhinav and Kadian, Abhishek and Al-Dahle, Ahmad and Letman, Aiesha and Mathur, Akhil and Schelten, Alan and Yang, Amy and Fan, Angela and others},
  journal={arXiv preprint arXiv:2407.21783},
  year={2024}
}

@inproceedings{lin2024rella,
  title={Rella: Retrieval-enhanced large language models for lifelong sequential behavior comprehension in recommendation},
  author={Lin, Jianghao and Shan, Rong and Zhu, Chenxu and Du, Kounianhua and Chen, Bo and Quan, Shigang and Tang, Ruiming and Yu, Yong and Zhang, Weinan},
  booktitle={Proceedings of the ACM on Web Conference 2024},
  pages={3497--3508},
  year={2024}
}

@inproceedings{DBLP:conf/iclr/LoshchilovH19,
  author       = {Ilya Loshchilov and
                  Frank Hutter},
  title        = {Decoupled Weight Decay Regularization},
  booktitle    = {7th International Conference on Learning Representations, {ICLR} 2019,
                  New Orleans, LA, USA, May 6-9, 2019},
  publisher    = {OpenReview.net},
  year         = {2019},
  url          = {https://openreview.net/forum?id=Bkg6RiCqY7},
  bibsource    = {dblp computer science bibliography, https://dblp.org}
}

@inproceedings{DBLP:conf/nips/VaswaniSPUJGKP17,
  author       = {Ashish Vaswani and
                  Noam Shazeer and
                  Niki Parmar and
                  Jakob Uszkoreit and
                  Llion Jones and
                  Aidan N. Gomez and
                  Lukasz Kaiser and
                  Illia Polosukhin},
  editor       = {Isabelle Guyon and
                  Ulrike von Luxburg and
                  Samy Bengio and
                  Hanna M. Wallach and
                  Rob Fergus and
                  S. V. N. Vishwanathan and
                  Roman Garnett},
  title        = {Attention is All you Need},
  booktitle    = {Advances in Neural Information Processing Systems 30: Annual Conference
                  on Neural Information Processing Systems 2017, December 4-9, 2017,
                  Long Beach, CA, {USA}},
  pages        = {5998--6008},
  year         = {2017},
  url          = {https://proceedings.neurips.cc/paper/2017/hash/3f5ee243547dee91fbd053c1c4a845aa-Abstract.html},
  bibsource    = {dblp computer science bibliography, https://dblp.org}
}

@article{bai2023qwen,
  title={Qwen technical report},
  author={Bai, Jinze and Bai, Shuai and Chu, Yunfei and Cui, Zeyu and Dang, Kai and Deng, Xiaodong and Fan, Yang and Ge, Wenbin and Han, Yu and Huang, Fei and others},
  journal={arXiv preprint arXiv:2309.16609},
  year={2023}
}

@article{zhao2023survey,
  title={A survey of large language models},
  author={Zhao, Wayne Xin and Zhou, Kun and Li, Junyi and Tang, Tianyi and Wang, Xiaolei and Hou, Yupeng and Min, Yingqian and Zhang, Beichen and Zhang, Junjie and Dong, Zican and others},
  journal={arXiv preprint arXiv:2303.18223},
  year={2023}
}

@article{wu2024survey,
  title={A survey on large language models for recommendation},
  author={Wu, Likang and Zheng, Zhi and Qiu, Zhaopeng and Wang, Hao and Gu, Hongchao and Shen, Tingjia and Qin, Chuan and Zhu, Chen and Zhu, Hengshu and Liu, Qi and others},
  journal={World Wide Web},
  volume={27},
  number={5},
  pages={60},
  year={2024},
  publisher={Springer}
}

@article{liu2024end,
  title={End-to-End Learnable Item Tokenization for Generative Recommendation},
  author={Liu, Enze and Zheng, Bowen and Ling, Cheng and Hu, Lantao and Li, Han and Zhao, Wayne Xin},
  journal={arXiv preprint arXiv:2409.05546},
  year={2024}
}

@inproceedings{wang2024learnable,
  title={Learnable item tokenization for generative recommendation},
  author={Wang, Wenjie and Bao, Honghui and Lin, Xinyu and Zhang, Jizhi and Li, Yongqi and Feng, Fuli and Ng, See-Kiong and Chua, Tat-Seng},
  booktitle={Proceedings of the 33rd ACM International Conference on Information and Knowledge Management},
  pages={2400--2409},
  year={2024}
}

@misc{wang2025msltokensneedtuning,
      title={MSL: Not All Tokens Are What You Need for Tuning LLM as a Recommender}, 
      author={Bohao Wang and Feng Liu and Jiawei Chen and Xingyu Lou and Changwang Zhang and Jun Wang and Yuegang Sun and Yan Feng and Chun Chen and Can Wang},
      year={2025},
      eprint={2504.04178},
      archivePrefix={arXiv},
      primaryClass={cs.IR},
      url={https://arxiv.org/abs/2504.04178}, 
}

@inproceedings{cui2024distillation,
  title={Distillation matters: empowering sequential recommenders to match the performance of large language models},
  author={Cui, Yu and Liu, Feng and Wang, Pengbo and Wang, Bohao and Tang, Heng and Wan, Yi and Wang, Jun and Chen, Jiawei},
  booktitle={Proceedings of the 18th ACM Conference on Recommender Systems},
  pages={507--517},
  year={2024}
}

@inproceedings{zheng2024adapting,
  title={Adapting large language models by integrating collaborative semantics for recommendation},
  author={Zheng, Bowen and Hou, Yupeng and Lu, Hongyu and Chen, Yu and Zhao, Wayne Xin and Chen, Ming and Wen, Ji-Rong},
  booktitle={2024 IEEE 40th International Conference on Data Engineering (ICDE)},
  pages={1435--1448},
  year={2024},
  organization={IEEE}
}

@article{rajput2023recommender,
  title={Recommender systems with generative retrieval},
  author={Rajput, Shashank and Mehta, Nikhil and Singh, Anima and Hulikal Keshavan, Raghunandan and Vu, Trung and Heldt, Lukasz and Hong, Lichan and Tay, Yi and Tran, Vinh and Samost, Jonah and others},
  journal={Advances in Neural Information Processing Systems},
  volume={36},
  pages={10299--10315},
  year={2023}
}

@inproceedings{ni2022sentence,
  title={Sentence-t5: Scalable sentence encoders from pre-trained text-to-text models},
  author={Ni, Jianmo and Abrego, Gustavo Hernandez and Constant, Noah and Ma, Ji and Hall, Keith and Cer, Daniel and Yang, Yinfei},
  booktitle={Findings of the association for computational linguistics: ACL 2022},
  pages={1864--1874},
  year={2022}
}

@article{cui2025hatllm,
  title={HatLLM: Hierarchical Attention Masking for Enhanced Collaborative Modeling in LLM-based Recommendation},
  author={Cui, Yu and Liu, Feng and Chen, Jiawei and Jin, Canghong and Lou, Xingyu and Zhang, Changwang and Wang, Jun and Sun, Yuegang and Wang, Can},
  journal={arXiv preprint arXiv:2510.10955},
  year={2025}
}

@inproceedings{zheng2024harnessing,
  title={Harnessing large language models for text-rich sequential recommendation},
  author={Zheng, Zhi and Chao, Wenshuo and Qiu, Zhaopeng and Zhu, Hengshu and Xiong, Hui},
  booktitle={Proceedings of the ACM Web Conference 2024},
  pages={3207--3216},
  year={2024}
}

@article{zhang2024text,
  title={Text-like encoding of collaborative information in large language models for recommendation},
  author={Zhang, Yang and Bao, Keqin and Yan, Ming and Wang, Wenjie and Feng, Fuli and He, Xiangnan},
  journal={arXiv preprint arXiv:2406.03210},
  year={2024}
}

@inproceedings{liu2025generative,
  title={Generative recommender with end-to-end learnable item tokenization},
  author={Liu, Enze and Zheng, Bowen and Ling, Cheng and Hu, Lantao and Li, Han and Zhao, Wayne Xin},
  booktitle={Proceedings of the 48th International ACM SIGIR Conference on Research and Development in Information Retrieval},
  pages={729--739},
  year={2025}
}

@inproceedings{lin2025recommendation,
  title={How do recommendation models amplify popularity bias? An analysis from the spectral perspective},
  author={Lin, Siyi and Gao, Chongming and Chen, Jiawei and Zhou, Sheng and Hu, Binbin and Feng, Yan and Chen, Chun and Wang, Can},
  booktitle={Proceedings of the Eighteenth ACM International Conference on Web Search and Data Mining},
  pages={659--668},
  year={2025}
}

@article{wang2025llm4dsr,
  title={Llm4dsr: Leveraging large language model for denoising sequential recommendation},
  author={Wang, Bohao and Liu, Feng and Zhang, Changwang and Chen, Jiawei and Wu, Yudi and Zhou, Sheng and Lou, Xingyu and Wang, Jun and Feng, Yan and Chen, Chun and others},
  journal={ACM Transactions on Information Systems},
  volume={44},
  number={1},
  pages={1--32},
  year={2025},
  publisher={ACM New York, NY}
}

@article{zhu2025llm,
  title={Llm as gnn: Graph vocabulary learning for text-attributed graph foundation models},
  author={Zhu, Xi and Xue, Haochen and Zhao, Ziwei and Xu, Wujiang and Huang, Jingyuan and Guo, Minghao and Wang, Qifan and Zhou, Kaixiong and Razzak, Imran and Zhang, Yongfeng},
  journal={arXiv preprint arXiv:2503.03313},
  year={2025}
}

@inproceedings{lin2024box,
  title={When box meets graph neural network in tag-aware recommendation},
  author={Lin, Fake and Zhao, Ziwei and Zhu, Xi and Zhang, Da and Shen, Shitian and Li, Xueying and Xu, Tong and Zhang, Suojuan and Chen, Enhong},
  booktitle={Proceedings of the 30th ACM SIGKDD Conference on Knowledge Discovery and Data Mining},
  pages={1770--1780},
  year={2024}
}

@article{zhu2025recommender,
  title={Recommender systems meet large language model agents: A survey},
  author={Zhu, Xi and Wang, Yu and Gao, Hang and Xu, Wujiang and Wang, Chen and Liu, Zhiwei and Wang, Kun and Jin, Mingyu and Pang, Linsey and Weng, Qingsong and others},
  journal={Foundations and Trends{\textregistered} in Privacy and Security},
  volume={7},
  number={4},
  pages={247--396},
  year={2025},
  publisher={Now Publishers, Inc.}
}

@inproceedings{wang2025unleashing,
  title={Unleashing the Power of Large Language Model for Denoising Recommendation},
  author={Wang, Shuyao and Zheng, Zhi and Sui, Yongduo and Xiong, Hui},
  booktitle={Proceedings of the ACM on Web Conference 2025},
  pages={252--263},
  year={2025}
}

@inproceedings{jiang2025beyond,
  title={Beyond Utility: Evaluating LLM as Recommender},
  author={Jiang, Chumeng and Wang, Jiayin and Ma, Weizhi and Clarke, Charles LA and Wang, Shuai and Wu, Chuhan and Zhang, Min},
  booktitle={Proceedings of the ACM on Web Conference 2025},
  pages={3850--3862},
  year={2025}
}

@article{wang2025beyond,
  title={Beyond Semantic Understanding: Preserving Collaborative Frequency Components in LLM-based Recommendation},
  author={Wang, Minhao and He, Yunhang and Xu, Cong and Zhu, Zhangchi and Zhang, Wei},
  journal={arXiv preprint arXiv:2508.10312},
  year={2025}
}

\appendix

\section{Baselines Details}
\label{sec:baseline}
\begin{itemize}
    \item \textbf{LightGCN}\cite{DBLP:conf/sigir/0001DWLZ020} A streamlined graph convolutional network for collaborative filtering, LightGCN focuses solely on neighborhood aggregation, omitting feature transformations and nonlinear activations to enhance efficiency and performance.
    \item \textbf{SASRec}\cite{DBLP:conf/icdm/KangM18} A self-attention-based sequential recommendation model that leverages the Transformer \cite{DBLP:conf/nips/VaswaniSPUJGKP17} architecture to capture users' dynamic interests by modeling the sequence of their historical interactions.
    \item \textbf{BERT4Rec}\cite{sun2019bert4rec} A bidirectional Transformer to train with a masking objective, allowing each token to attend to both left and right context.
    \item \textbf{AlphaRec}\cite{sheng2024language} One of the latest recommender models that leverages textual metadata (e.g. item titles) by mapping their language representations to collaborative space for recommendation.
    \item \textbf{RLMRec}\cite{ren2024representation} The first LLM–CF embedding alignment framework that unifies collaborative relational embeddings and language model–derived semantic representations into a shared space via contrastive and generative alignment techniques.
    \item \textbf{TIGER} \cite{rajput2023recommender} The first work that leverages RQ‑VAE to encode items into hierarchical Semantic IDs and trains a Transformer to autoregressively generate these IDs as retrieval targets.
    \item \textbf{LETTER} \cite{wang2024learnable} A method that learns semantic IDs by integrating collaborative signals into RQ-VAE item tokenization for LLM‑based generative recommendation.
    \item \textbf{TallRec} \cite{DBLP:journals/corr/abs-2305-00447} An efficient tuning framework that aligns large language models with recommendation tasks through instruction tuning and LoRA-based fine-tuning, enhancing performance even with limited training data.
    \item 
    \textbf{Llara}\cite{liao2024llara}A novel hybrid prompting method that integrates ID-based item embeddings learned by traditional recommendation models with textual item features
    \item \textbf{Collm}\cite{DBLP:journals/corr/abs-2310-19488} A method that integrates collaborative filtering embeddings into large language models by mapping external collaborative information into the LLM's input space, improving recommendation quality without modifying the LLM itself.
    \item \textbf{MSL}\cite{wang2025msltokensneedtuning} An approach masks invalid tokens during loss computation so that the loss focuses on item title tokens and avoids penalizing spurious or fictitious text generations.
\end{itemize}
\section{Experimental Settings}
\label{sec:exp_setting}

\subsection{Implementation Details}
We use LLama3.2-3B\cite{dubey2024llama} as our backbone and SASRec\cite{DBLP:conf/icdm/KangM18} as our base recommendation model. For training efficiency, we employ LoRA\cite{DBLP:conf/iclr/HuSWALWWC22} to fine-tune the LLMs, a technique that has been  proven to be highly effective for training LLMs. All experiments are conducted on an RTX A6000. Similar to previous works\cite{lin2024rella,DBLP:journals/corr/abs-2310-19488}, the LoRA parameters are set with a rank of 8, alpha of 16, and a dropout rate of 0.05. For traditional ID-based methods, the embedding dimension is uniformly set to 64, following the full-shot setting. For large model-based methods, we use AdamW\cite{DBLP:conf/iclr/LoshchilovH19} as the optimizer, with a learning rate that follows a linear scheduler, initializing at 1e-4. For all methods related to LLMs, we set the maximum number of epochs to 10, with evaluation every 2 epochs. We also employ early stopping, halting training if there is no improvement after 2 consecutive evaluations. We set the historical item sequence length as 10 and the maximum token length as 2048 to ensure efficient training. 
\subsection{Evaluation Metrics}
To assess performance, we use Hit Ratio (HR) and normalized Discounted Cumulative Gain (NDCG) with 
k in [5, 10], where HR measures whether the ground-truth item is hit within the top-k results and NDCG evaluates the quality of its ranking position.

\subsection{Inference Setting}
During LLM inference, we employ a constrained prefix tree to ensure that the generated text strictly corresponds to items within our candidate list. 
\section{Derivation of Auxiliary KL Loss}
\label{form:aux_detail}

\begin{equation*}
\begin{aligned}
\mathcal{L}_{\text{aux}}
&= (1-\alpha)\,\mathcal{L}_{\text{NTP}}
\;+\;
\alpha \sum_{j=1}^{|t|}
\mathrm{KL}\!\left(p_u(\cdot \mid y_{<j})
\,\|\, 
P(\cdot \mid x_u, y_{<j}; \theta)\right). \\
&=(1-\alpha)\,\mathcal{L}_{\mathrm{NTP}}
  + \alpha \sum_{j=1}^{|t|}
  \mathrm{KL}\!\big(p_j \,\|\, P_j\big),
\\
&=(1-\alpha)\,\mathcal{L}_{\mathrm{NTP}}
  + \alpha \sum_{j=1}^{|t|}
    \sum_{v\in\mathcal V} p_j(v)\,
    \big[\log p_j(v) - \log P_j(v)\big],
\\
&= -\sum_{j=1}^{|t|}\sum_{v\in\mathcal V}
\underbrace{\big[(1-\alpha)\mathbf{1}_{v=y_j}+\alpha\,p_j(v)\big]}_{\displaystyle \tilde{y}_j(v)}
\log P_j(v)
\;+\; \text{const},\\
&= - \sum_{j=1}^{|t|}\sum_{v\in\mathcal V}
     \tilde{y}_j(v)\,\log P_j(v)
   + \text{const}.
\end{aligned}
\end{equation*}

\section{Gradient of Auxiliary KL}
\label{sec:aux_grad}
We start from the auxiliary KL loss at a single position $j$:
\[
L_j^{aux} = - \sum_{v\in\mathcal{V}} \tilde{y}_j(v)\,\log p_j(v),
\]
where $p_j(v) = \frac{\exp(z_j(v))}{\sum_{u\in\mathcal{V}} \exp(z_j(u))}$ is the softmax probability and $z_j(v)$ denotes the logit of token $v$ at position $j$.

Expanding $\log p_j(v)$ gives
\[
\log p_j(v) = z_j(v) - \log \sum_{u\in\mathcal{V}} \exp(z_j(u)).
\]

Plugging this into the loss:
\[
L_j^{aux} = - \sum_{v} \tilde{y}_j(v) z_j(v)
+ \Big(\sum_{v}\tilde{y}_j(v)\Big)\,\log \sum_{u} \exp(z_j(u)).
\]

Taking the derivative with respect to $z_j(k)$:
\[
\frac{\partial L_j^{aux}}{\partial z_j(k)}
= -\,\tilde{y}_j(k)
+ \Big(\sum_{v}\tilde{y}_j(v)\Big)\,\frac{\exp(z_j(k))}{\sum_{u}\exp(z_j(u))}.
\]

Since $\sum_v \tilde{y}_j(v)=1$, this simplifies to
\[
\;\frac{\partial L_j}{\partial z_j(k)} = p_j(k) - \tilde{y}_j(k).\;
\]
\section{Gradient of Soft NTP}
\label{sec:softNTP_grad}
Similarly, we start from the soft-NTP loss at a single position $j$:
\[
L^{\text{soft-NTP}}_j = - \log \Bigg(\sum_{v\in\mathcal{V}} \tilde{y}_j(v)\,p_j(v)\Bigg),
\]
where $p_j(v) = \frac{\exp(z_j(v))}{\sum_{u\in\mathcal{V}} \exp(z_j(u))}$ is the softmax probability and $z_j(v)$ denotes the logit of token $v$ at position $j$.

Let
\[
S_j = \sum_{v\in\mathcal{V}} \tilde{y}_j(v)\,p_j(v),
\]
so that $L^{\text{soft-NTP}}_j = -\log S_j$. Taking the derivative with respect to $z_j(k)$ gives
\[
\frac{\partial L^{\text{soft-NTP}}_j}{\partial z_j(k)} 
= -\frac{1}{S_j}\,\frac{\partial S_j}{\partial z_j(k)}.
\]
Since
\[
\frac{\partial S_j}{\partial z_j(k)}
= \sum_{v} \tilde{y}_j(v)\,\frac{\partial p_j(v)}{\partial z_j(k)}
= \tilde{y}_j(k)\,p_j(k) - p_j(k)\sum_{v}\tilde{y}_j(v)\,p_j(v),
\]
we obtain
\[
\frac{\partial S_j}{\partial z_j(k)} = p_j(k)\big(\tilde{y}_j(k) - S_j\big).
\]
Plugging this back into the derivative of the loss:
\[
\frac{\partial L^{\text{soft-NTP}}_j}{\partial z_j(k)}
= -\frac{1}{S_j}\,p_j(k)\big(\tilde{y}_j(k)-S_j\big)
= p_j(k) - \frac{\tilde{y}_j(k)\,p_j(k)}{S_j}.
\]

Defining
\[
q_j(k) = \frac{\tilde{y}_j(k)\,p_j(k)}{\sum_{u}\tilde{y}_j(u)\,p_j(u)} 
= \frac{\tilde{y}_j(k)\,p_j(k)}{S_j},
\]
we finally arrive at
\[
\;\frac{\partial L^{\text{soft-NTP }}_j}{\partial z_j(k)} = p_j(k) - q_j(k).\;
\]

\end{document}